\documentclass[11pt,a4paper]{article}
\usepackage{jcappub}

\newcommand{\de}{{\mathrm d}}
\newcommand{\om}{{\Omega_m}}
\newcommand{\ob}{{\Omega_b}}
\newcommand{\odm}{{\Omega_\mathrm{DM}}}

\newcommand{\ho}{{H_0}}
\newcommand{\lcdm}{$\Lambda$CDM}

\newcommand{\sbo}{{\sigma_{B,0}}}
\newcommand{\am}{{\langle\alpha-1\rangle}}

\title{Magnification bias as a novel probe for primordial magnetic fields}

\author[a]{S. Camera,}
\author[b,c]{C. Fedeli}
\author[d,b,e]{and L. Moscardini}
\affiliation[a]{CENTRA, Instituto Superior T\'ecnico, Universidade de Lisboa,\\Av. Rovisco Pais 1, 1049-001 Lisboa, Portugal}
\affiliation[b]{INAF - Osservatorio Astronomico di Bologna,\\Via Ranzani 1, 40127 Bologna, Italy}
\affiliation[c]{Department of Astronomy, University of Florida,\\211 Bryant Space Science Center, Gainesville, FL 32611, USA}
\affiliation[d]{Dipartimento di Fisica e Astronomia, Universit\`a di Bologna,\\Viale Berti Pichat 6/2, 40127 Bologna, Italy}
\affiliation[e]{INFN - Sezione di Bologna,\\Viale Berti Pichat 6/2, 40127 Bologna, Italy}
\emailAdd{stefano.camera@tecnico.ulisboa.pt}
\emailAdd{cosimo.fedeli@oabo.inaf.it}
\emailAdd{lauro.moscardini@unibo.it}

\abstract{In this paper we investigate magnetic fields generated in the early Universe. These fields are important candidates at explaining the origin of astrophysical magnetism observed in galaxies and galaxy clusters, whose genesis is still by and large unclear. Compared to the standard inflationary power spectrum, intermediate to small scales would experience further substantial matter clustering, were a cosmological magnetic field present prior to recombination. As a consequence, the bias and redshift distribution of galaxies would also be modified. Hitherto, primordial magnetic fields (PMFs) have been tested and constrained with a number of cosmological observables, e.g.\ the cosmic microwave background radiation, galaxy clustering and, more recently, weak gravitational lensing. Here, we explore the constraining potential of the density fluctuation bias induced by gravitational lensing magnification onto the galaxy-galaxy angular power spectrum. Such an effect is known as \textit{magnification bias}. 
Compared to the usual galaxy clustering approach, magnification bias helps in lifting the pathological degeneracy present amongst power spectrum normalisation and galaxy bias. This is because magnification bias cross-correlates galaxy number density fluctuations of nearby objects with weak lensing distortions of high-redshift sources. Thus, it takes advantage of the gravitational deflection of light, which is insensitive to galaxy bias but powerful in constraining the density fluctuation amplitude. To scrutinise the potentiality of this method, we adopt a deep and wide-field spectroscopic galaxy survey. We show that magnification bias does contain important information on primordial magnetism, which will be useful in combination with galaxy clustering and shear. We find we shall be able to rule out at $95.4\%$ CL amplitudes of PMFs larger than $5\times10^{-4}\,\mathrm{nG}$ for values of the PMF power spectral index $n_B\sim0$.}


\arxivnumber{}

\begin{document}

\maketitle

\section{Introduction}
The presence of weak magnetic fields in the very early Universe (henceforth primordial magnetic fields, or PMFs) has been subject of investigation for several decades now \citep{1965JETP...21..656Z,Thorne:1967zz,1972A&A....19..189R}. A number of theoretical models predict the formation of such magnetic fields in the course of inflation \citep{Turner:1987bw,1992ApJ...391L...1R} or phase transitions \citep{Quashnock:1988vs}; the amplification of PMFs during the process of structure formation provides a possible explanation for the magnetic fields that are observed nowadays in galaxies and galaxy clusters \citep{Widrow:2002ud,Schober:2012nm}. One piece of evidence in support of this scenario would be the detection of magnetic fields on cosmological scales. However, as of today, Faraday rotation measures have only been able to set upper bounds on the strength of cosmological magnetic fields, at the level of $0.1-1\,\mathrm{nG}$ on a $\mathrm{Mpc}$ scale \citep{1970ApL.....6..169B,1990ApJ...360....1V,1994RPPh...57..325K,Kolatt:1997xu,Blasi:1999hu} (see also \citep{Xu:2005rb}). Additional upper bounds at the level (extrapolated at present) of $\lesssim 1$ nG have also been derived from observations of the cosmic microwave background (CMB) radiation \citep{Barrow:1997mj,Subramanian:1998fn,Jedamzik:1999bm,Subramanian:2002nh,Yamazaki:2011eu,Yamazaki:2010nf,Yamazaki:2010jw,Yadav:2012uz,Ade:2013zuv} and the abundance of elements in the Universe \citep{Cheng:1993kz,Kernan:1995bz,Cheng:1996yi,Grasso:1996kk,Kawasaki:2012va}. Conversely, lower bounds have been put by employing for instance gamma ray data \citep{Essey:2010nd}.

If PMFs indeed existed, their presence would have altered the statistics of inflationary density fluctuations which subsequently originated, through gravitational instability, the observed large-scale cosmic structure. To be more specific, PMFs would have been able to produce and amplify density fluctuations in the baryonic fluid through the Lorentz force. After recombination these fluctuations would have grown freely and imprinted similar fluctuations on the dark matter fluid, to which baryons are gravitationally coupled. The final result would be a distortion in the linear power spectrum of density fluctuations, dependent on the strength and statistical distribution of PMFs \citep{1978ApJ...224..337W,Kim:1996,Subramanian:1997gi}. This opens an interesting new window to probe early magnetism, based on its effects on the large-scale structure formation. As of now, applications in this sense include the formation of voids \citep{deAraujo:1997dz}, cosmic re-ionisation \citep{Sethi:2005,Tashiro:2006uv}, and 
$21$ cm fluctuations from neutral hydrogen \citep{Tashiro:2005ua}.

Such an intriguing hypothesis has been tested by various works of different groups, who have proven this on galaxy and galaxy cluster scales in full cosmological, magneto-hydrodynamical simulations \citep[e.g.][]{2009IAUS..259..667A,Arshakian:2009dr,2010MNRAS.403..453D,Kahniashvili:2012uj,Pakmor:2013rqa}. However, those simulations do not introduce any modification to the primordial spectrum of perturbations, but rather evolve the purely inflationary spectrum in the presence of PMFs. Conversely, in a previous paper by our group \citep{Fedeli:2012rr}, we adopted a complementary approach especially devoted to highlight and understand the effects of PMFs from a cosmological viewpoint. In particular, we used for the first time a self-consistent approach inspired to the halo model of structure formation \citep{Cooray:2002dia} to describe the non-linear evolution of structures.\footnote{See also \citep{Pandey:2012kk} for a similar study that however does not include a full treatment of non-linearities.} We then used this to show that future cosmic shear datasets, on the model of the \textit{photometric} catalogue of the ESA Euclid\footnote{\texttt {http://www.euclid-ec.org}} survey \citep{EditorialTeam:2011mu}, will be able to improve substantially the current bounds on the amplitude of PMFs. In an effort to complement this previous study, we here investigated how PMFs impact the so-called magnification bias. For consistency, we forecast constraints on PMFs given by magnification bias measured in a Euclid-like \textit{spectroscopic} catalogue.

As a matter of fact, another explanation for the presence of magnetic fields in galaxies and galaxy clusters may come from astrophysical mechanisms, such as supernov\ae, active galactic nuclei, galactic outflows or even battery effects \cite{1950ZNatA...5...65B,1996ARA&A..34..155B,1972SvA....15..714V,Neronov:1900zz,Tavecchio:2010mk,2011MNRAS.414.3566T,Neronov:2013zka}. However, neither a cosmological origin nor the astrophysical interpretation have hitherto been thoroughly confirmed by observations. Therefore, we find it worth investigating to analyse the effects of PMFs, particularly in the light of the new generation of wide-field experiments that will soon provide us with accurate data on the large-scale cosmic structure and its evolution.

Here, we focus on magnification bias, a well-known effect \cite[e.g.][]{LoVerde:2007ke,Kostelecky:2008iz}. It consists of a modulation of galaxy counts due to the cross-correlation between foreground and background sources induced by weak lensing magnification. This happens because intervening matter along the line of sight gravitationally lenses high-redshift sources. Thus, on the one hand the observed galaxies per unit area on the sky are fewer (because of the stretching of the apparent intergalaxy spacing). On the other hand fainter galaxies, which would be below the detectability threshold, become visible, thus leading to a larger number density. The net effect depends on the number count slope and leads to a correction to the intrinsic galaxy density, i.e.
\begin{equation}
\delta_g\to\delta_g+\delta_\mu\;,
\end{equation}
where $g$ stands for galaxy number counts and $\mu$ labels magnification.

The main idea behind this endeavour is to be able eventually to combine all the various pieces of information provided by oncoming large-scale surveys such as the Euclid satellite into a thorough set of constraints on PMFs. This will necessarily also include the redshift-space galaxy correlation function, which has already been addressed by references~\citep{2003JApA...24...51G,Sethi:2003vp}, but with neither a direct reference to future observational efforts nor a self-consistent approach to non-linear structure formation. In spite of the fact that the strongest signal is expected to come from cosmic shear and galaxy clustering, in this paper we show that the magnification bias, for all being a second-order effect, can by itself also put competitive bounds on the amplitude of PMFs.

The paper is structured as follows: section~\ref{sec:magnification} introduces the cardinal observable of this work, i.e.\ the magnification bias angular power spectrum; sect.~\ref{sec:pmf} briefly describes the theory of cosmological PMFs; sect.~\ref{sec:clustering} reviews the main modifications induced by PMFs on clustering of haloes; in sect.~\ref{sec:survey} we define the reference survey and its specifications; in sect.~\ref{sec:forecasts} we present the main results of our work, viz.\ forecast constraints on PMF parameters, and we also discuss the assumptions we make; and eventually sect.~\ref{sec:conclusions} draws the major conclusions.

Throughout the paper, we assume as reference model a flat \lcdm\ Universe, whose background evolution is determined by the Hubble rate $H\equiv\de\ln a/\de t$, where $a(t)$ is the scale factor at cosmic time $t$. The Hubble rate can be expressed as
\begin{equation}
H(z)=\ho\sqrt{\om\left(1+z\right)^3+1-\om}\;,\label{eq:hubble}
\end{equation}
where $z=1/a-1$ is the redshift, $\ho=100h\,\mathrm{km\,s^{-1}\,Mpc^{-1}}$ is the Hubble constant in units of dimensionless parameter $h$, and $\om=\odm+\ob$ is the total matter fraction, given $\odm$ and $\ob$ respectively the dark matter and baryon densities in units of the critical density. The Universe's expansion history also defines the radial comoving distance $\chi(z)$ from an observer to an object located at redshift $z$ via $\de\chi=\de z/H$. Note that we use units such that the speed of light $c=1$.

\section{Magnification bias}\label{sec:magnification}
The cardinal observable that we studied in this paper is one of the members of the vast family of gravitational lensing phenomena. Light rays are known to be deflected by the cosmic large-scale structure along the line of sight, which systematically introduces distortions in the observed images of distant sources. As a consequence, sources behind a lens are magnified in size, whilst their surface brightness is conserved. This leads to an increase in the total observed luminosity of a source. From an observational point of view one can detect the effects of this magnification by cross-correlating two disjoint redshift distributions of sources: the low-redshift `lenses', which act as a foreground that magnifies the background, high-redshift `sources'. This cosmic magnification effect was first detected by ref.~\citep{Scranton:2005ci} by cross-correlating low-redshift galaxies from the Sloan Digital Sky Survey (SDSS) with SDSS quasars. More recently, ref.~\citep{2009A&A...507..683H} has detected the effect in samples of normal galaxies in the Canada-France-Hawaii-Telescope Legacy Survey, whilst for instance ref.~\citep{Menard:2009yb} has built on the SDSS analysis by constraining galaxy-mass and galaxy-dust correlation functions.

The effect can be formally described as follows. At a position $\boldsymbol\theta$ on the celestial sphere, we can relate the behaviour of unlensed sources with number density $\overline{N_s}(<m)$, where $m$ is their apparent magnitude, to that of lensed sources with number density $N_s(<m,\boldsymbol\theta)$. It must be noted that there are two competing effects: \textit{i)} the flux increases due to magnification of distant faint sources and thus augments the number density of observed images above a certain magnitude threshold; \textit{ii)} counteracting this, the dilution of the number density due to the stretching of the solid angle caused by lensing. If the source fluxes have a distribution with a power-law slope given by
\begin{equation}
\alpha(m)=\frac{5}{2}\frac{\partial}{\partial m}\mathrm{Log}\left[\overline{N_s}(<m)\right]\;,\label{eq:alpha}
\end{equation}
one can obtain \citep{Bartelmann:1999yn,Bartelmann:2010fz}
\begin{equation}
N_s(<m,\boldsymbol\theta)dm=\mu^{\alpha(m)-1}\overline{N_s}(<m)dm\;,\label{eq:magnification}
\end{equation}
where the magnification $\mu$ is
\begin{equation}\label{eq:mag}
\mu=\frac{1}{\left|(1-\kappa)^2-\left|\gamma\right|^{2}\right|}\;.
\end{equation}
Here, $\kappa$ is the convergence and $\gamma$ the (complex) shear; they are the two fundamental, first-order weak lensing distortions. In the weak lensing r\'egime, it is known that $\mu\simeq1+2\kappa$, whence it is clear that magnification is closely related to convergence. In turn, the convergence is a weighted estimator of the matter density fluctuations along the line-of-sight direction \citep{Bartelmann:1999yn}, viz.
\begin{equation}
\kappa(\boldsymbol\theta)=\int_0^{+\infty}\!\!\de\chi\,W^{\kappa_s}(\chi)\chi\frac{\delta(\chi\boldsymbol\theta,\chi)}{a(\chi)}\;,
\end{equation}
with
\begin{equation}
W^{\kappa_s}(\chi)=\frac{3}{2}\ho^2\om\int_\chi^{+\infty}\!\!\de\chi'\,\frac{\de N_s}{\de\chi'}\frac{\chi'-\chi}{\chi'}\label{eq:Ws}
\end{equation}
being the weak lensing weight function and $(\de N_s/\de z)\de z=(\de N_s/\de\chi)\de\chi$ the source redshift distribution.\footnote{It is worth noting that the integral in eq.~\eqref{eq:Ws} actually extends up to the horizon comoving distance $\chi_H$. However, the upper integration limit may be formally widened to infinity, since the source redshift distribution has a depth due to experimental set ups which vanishes well before $\chi_H$.}

Because of the magnification bias effect, we can infer cosmological information by cross-correlating foreground and background objects, thus investigating how clustered background sources appear around foreground galaxies compared to a random distribution. To do so, we introduce the magnification bias angular power spectrum \citep{Bartelmann:1999yn,Kostelecky:2008iz}
\begin{equation}
C^{\mu_sg_l}(\ell)=2\am\int\!\!\de\chi\,\frac{W^{\kappa_s}(\chi)W^{g_l}(\chi)}{\chi^2}P^\delta\!\!\left(\frac{\ell}{\chi},\chi\right)\;,\label{eq:C^mg}
\end{equation}
where $W^{g_l}(\chi)$ is the weight function for the redshift distribution of the lenses $\de N_l/\de z(z)$, i.e.
\begin{equation}
W^{g_l}\left[\chi(z)\right]=H(z)b_l(z)\frac{\de N_l}{\de z}(z)\;,\label{eq:Wl}
\end{equation}
with $b_l(z)$ the bias of the galaxy population acting as lenses. We make use of Limber's approximation and fix $k=\ell/\chi$ \citep{1953ApJ...117..134L,Kaiser:1991qi}. As a matter of fact, Limber's approximation tends to underestimate the power at small angular multipoles, compared to the full spherical expansion \cite[e.g.][]{Kitching:2010wa}. Nonetheless, PMFs alter the galaxy clustering power spectrum at intermediate to small scales. In other words, the effects we are looking for kick in at multipoles much larger than those where Limber's approximation does not hold. Thus, we can safely adopt it in our analysis.

The pre-factor in eq.~\eqref{eq:C^mg}, $\alpha-1$, comes from the weak lensing r\'egime, where $\mu\ll1$ and eq.~\eqref{eq:magnification} reduces to $N_s=(\alpha-1)\overline{N_s}+\mathcal O(\mu^2)$. The ensemble average, $\langle\ldots\rangle$, is instead a consequence of the fact that one usually considers sources over a given magnitude range, where $\alpha(m)$ may vary. Therefore, it is more correct to define \citep{Scranton:2005ci}
\begin{equation}
\am=\left[\int\!\!\de m\,N(<m)\right]^{-1}\int\!\!\de m\,N(<m)\left[\alpha(m)-1\right]\;.\label{eq:am1}
\end{equation}
Clearly, magnification bias occurs only when $\alpha\ne1$.

\section{Primordial magnetic fields}\label{sec:pmf}
If the astrophysical magnetic fields that are measured in the large-scale structure are indeed the result of the amplification of PMFs, then the latter must have enhanced baryonic density fluctuations through the Lorentz force. This increment would then propagate to the dark matter density fluctuations through gravitational coupling. As it turns out, the general effect of PMFs is to enhance the matter  power on small scales, with the specific details depending on their statistical properties \citep{Kim:1996,Subramanian:1997gi}. The three-dimensional power spectrum $P_B(k,t)$ of PMFs, can be defined via \citep{Kraichnan:1967}
\begin{equation}
\left\langle \hat B_i(\mathbf k, t) \hat B_j(\mathbf k', t) \right \rangle = \frac{1}{2}(2\pi)^3 \delta_\mathrm{D}(\mathbf k - \mathbf k') \left( \delta_{ij} -\frac{k_ik_j}{k^2} \right) P_B(k,t)\;,
\end{equation}
where $\hat{\mathbf B}(\mathbf k, t)$ is the Fourier transform of the magnetic field vector at cosmic time $t$ and, as before, angled brackets denote ensemble averaging. It is customary to parameterise the PMF power spectrum as a simple power-law as a function of scale, $P_B(k,t) = A_B(t)k^{n_B}$. This power-law is then cut off at small scales, with the cut-off scale $k_\mathrm{max}$ interpreted as the wavenumber above which PMFs dissipate radiatively around the time of recombination \citep{Jedamzik:1996wp,Subramanian:1997gi,Mack:2001gc}. However, the magnetic Jeans scale $k_B$, beyond which magnetically induced density perturbations stop growing because magnetic pressure gradients counteract gravity \citep{Kim:1996,Subramanian:1997gi,Sethi:2005}, is usually much smaller than $k_\mathrm{max}$. This means that, for practical applications, the magnetically induced part of the matter power spectrum is effectively truncated at $k_B$.

Similarly to what happens for density fluctuations, the amplitude of the PMF power spectrum can also be expressed in terms of the variance of magnetic fields smoothed on some comoving scale $\lambda$. Specifically, by assuming a sharp filtering in Fourier space one obtains
\begin{equation}
\sigma_B^2(\lambda,t) \equiv \frac{1}{2\pi^2}\int_0^{2\pi/\lambda}\de k\,k^2P_B(k, t) = \frac{3}{3+n_B}A_B(t)\frac{(2\pi)^{1+n_B}}{\lambda^{3+n_B}}\;.
\end{equation}
On scales much larger than $k_B$ the generation of PMFs via magnetic induction by the baryonic velocity fields can be neglected, and the amplitude of magnetic fields decreases with cosmic expansion simply as $\propto a(t)^{-2}$. This allows one to extrapolate the PMF mean amplitude to its present value, $\sigma_{B,0}(\lambda)$, which is customarily used to set the PMF strength. From now on, we shall often use simply $\sigma_{B,0}$ as a shorthand notation for $\sigma_{B,0}(\lambda=1\,\mathrm{Mpc})$.

In ref.~\citep{Fedeli:2012rr} we summarised the details of how the small-scale increment in matter power due to the presence of PMFs can be computed for given magnetic amplitude and spectral slope. Here, we just collect the main features, and refer the reader to our previous paper for additional information. As it happens, the amplitude of the magnetically induced matter power spectrum grows as $\propto \sigma^4_{B,0}(\lambda)$, whilst its peak gets shifted to smaller and smaller scales as the magnetic spectral index decreases. This behaviour implies that the best constraints on the amplitude of PMFs from the large-scale structure can be obtained for large values of $n_B$. However, positive values of the magnetic spectral index are currently not observationally favoured \citep{Caprini:2001nb,Paoletti:2010rx,Kahniashvili:2012dy,Ade:2013zuv}, and we shall thence focus on $n_B\le0$. Furthermore, the growth rate for the magnetically induced matter power spectrum tends to the standard growth rate at late times, whilst it detaches from it as one gets closer to the recombination time. The reason is that PMFs can act only on baryons, and before recombination baryonic density fluctuations cannot grow due to the strong coupling with photons.

\section{Effects of primordial magnetic fields on magnification bias}\label{sec:clustering}
The presence of PMFs affects the magnification bias angular power spectrum in a twofold way. First, through the linear matter power spectrum, which gets enhanced on small scales as described in the previous section. Second, the enhancement in matter clustering that PMFs generate on top of the inflationary power spectrum has the generic effect of producing a larger mass variance for small density perturbations. This impacts the halo mass function and linear bias, both of which are ingredients of the theoretical computation of the magnification bias power spectrum. Again, we summarise below the important points, whilst further details can be found in ref.~\citep{Fedeli:2012rr}.

\subsection{Halo mass function and bias}\label{ssec:bias}
We employ the Sheth and Tormen mass function \citep{Sheth:2001dp} and the Sheth, Mo and Tormen
bias prescription \citep{Sheth:1999su}. These are physically motivated formalisms (particularly, by the ellipsoidal collapse model), we can therefore argue them to remain acceptably valid in cosmologies including primordial magnetism, as long as the mass variance is properly computed. The effect of PMFs on the halo mass function is rather complex. On galaxy cluster scales ($M\sim 10^{15}\,h^{-1}\,M_\odot$) the mass function is virtually unchanged, because of the fact that PMFs give a contribution to the matter power only at relatively small scales. On galaxy scales ($M\sim10^{12}\,h^{-1}\,M_\odot$) the halo abundance gets increased due to the presence of PMFs, because the mass variance increases and it also gets steeper with mass --- that is to say, the term $|\mathrm d\sigma_M/\mathrm d\log M|$ in the mass function becomes larger. However, on even smaller scales there is a trend reversal and the mass function decreases, because below the magnetic Jeans scale the variance flattens out.

The halo bias shows a behaviour that is consistent with that of the mass function. Specifically, the introduction of PMFs leaves the bias unchanged at large masses, whereas it generates a substantial bias reduction at intermediate and small masses. This is to be expected as more abundant objects are also less biased. As a final remark, we would like to emphasise that a more accurate modelling of the PMF effects on the small-scale power spectrum would require sophisticated numerical simulations over a wide range of scales. However, we argue that our approach is nonetheless self-consistent and theoretically motivated, since it only requires that the mass function and the bias are universal --- at least to some extent. This is indeed what happens in the case of dynamical dark energy \citep[e.g.][]{Grossi:2008xh,Baldi:2008ay,DeBoni:2010nz} and massive neutrinos \citep[e.g.][]{Marulli:2011he,Castorina:2013wga}. Eventually, we want to anticipate that most part of the magnification bias constraining power with respect to PMFs in facts comes from mildly large to intermediate scales, as will be clear in the following.

\subsection{Matter power spectrum}\label{ssec:powerspectrum}
Calculation of the magnification bias power spectrum shown in eq.~\eqref{eq:Wl} requires knowledge of the fully non-linear matter power spectrum. We estimate this by using the halo model, a semi-analytic framework that allows for a modelling of the non-linear clustering of matter using physically motivated ingredients \cite[see][for additional details]{Ma:2000ik,Seljak:2000gq,Cooray:2002dia}. To be more specific, the matter power spectrum can be written as the sum of a $1$-halo term, which dominates at small scales and depends on the average internal structure of dark matter haloes, and a $2$-halo term, which dominates on large scales and depends on the mutual clustering of individual haloes. We assumed that the dark matter density run can be described by a NFW \citep{Navarro:1996gj} profile for all cosmological models considered in this work, with concentrations given by the relation of Dolag et al.~\citep{Dolag:2003ui}. Conversely, we included the modifications to the halo abundance and bias produced by PMFs as described in the previous section.

\begin{figure}
\centering
\includegraphics[width=0.75\textwidth]{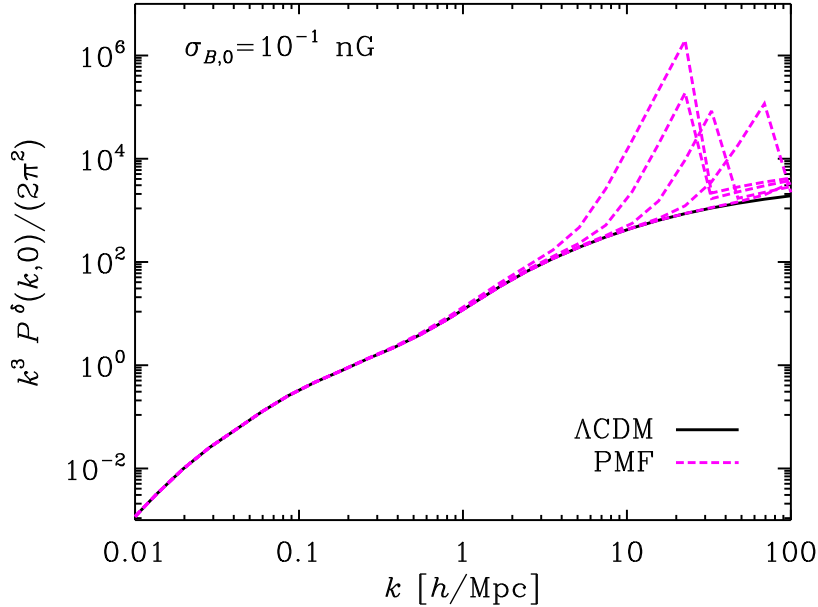}
\caption{Present-day matter power spectrum in the presence of PMFs (dashed, magenta curves) for $\sbo=10^{-1}\,\mathrm{nG}$ and various magnetic spectral indices, compared to the standard \lcdm\ prediction (solid, black line). Values of $n_B$ increase from $-2.9$ (bottom curve) to $0$ (top curve).}\label{fig:Pk-PMF}
\end{figure}

As already shown and discussed \citep{Fedeli:2012rr}, the inclusion of PMFs increases the non-linear matter power spectrum in two ways: directly, through the enhancement in matter clustering that is imprinted at small scales on the underlying linear matter power spectrum; and indirectly, through the increased mass variance, which permits the $1$-halo term to become important at larger scales as compared to a standard cosmology without PMFs. To better visualise it, fig.~\ref{fig:Pk-PMF} depicts the present-day matter power spectrum $P^\delta(k,0)$ both in the reference \lcdm\ cosmology (solid, black curve) and in the presence of PMFs (magenta, dashed lines). We compute the latter with $\sbo=10^{-1}\,\mathrm{nG}$ --- although such a large value is nowadays observationally disfavoured --- for the illustrative purpose of enhancing the PMF features and thus helping clarifying their effects onto the clustering of matter. This modified non-linear power spectrum, altogether with the modified halo mass function and bias, can now be inserted back into eq.~\eqref{eq:Wl} in order to forecast constraints on the amplitude $\sigma_{B,0}(\lambda)$ of PMFs. This requires the specification of a survey configuration.

\section{The reference survey}\label{sec:survey}
\begin{figure}
\centering
\includegraphics[width=0.5\textwidth]{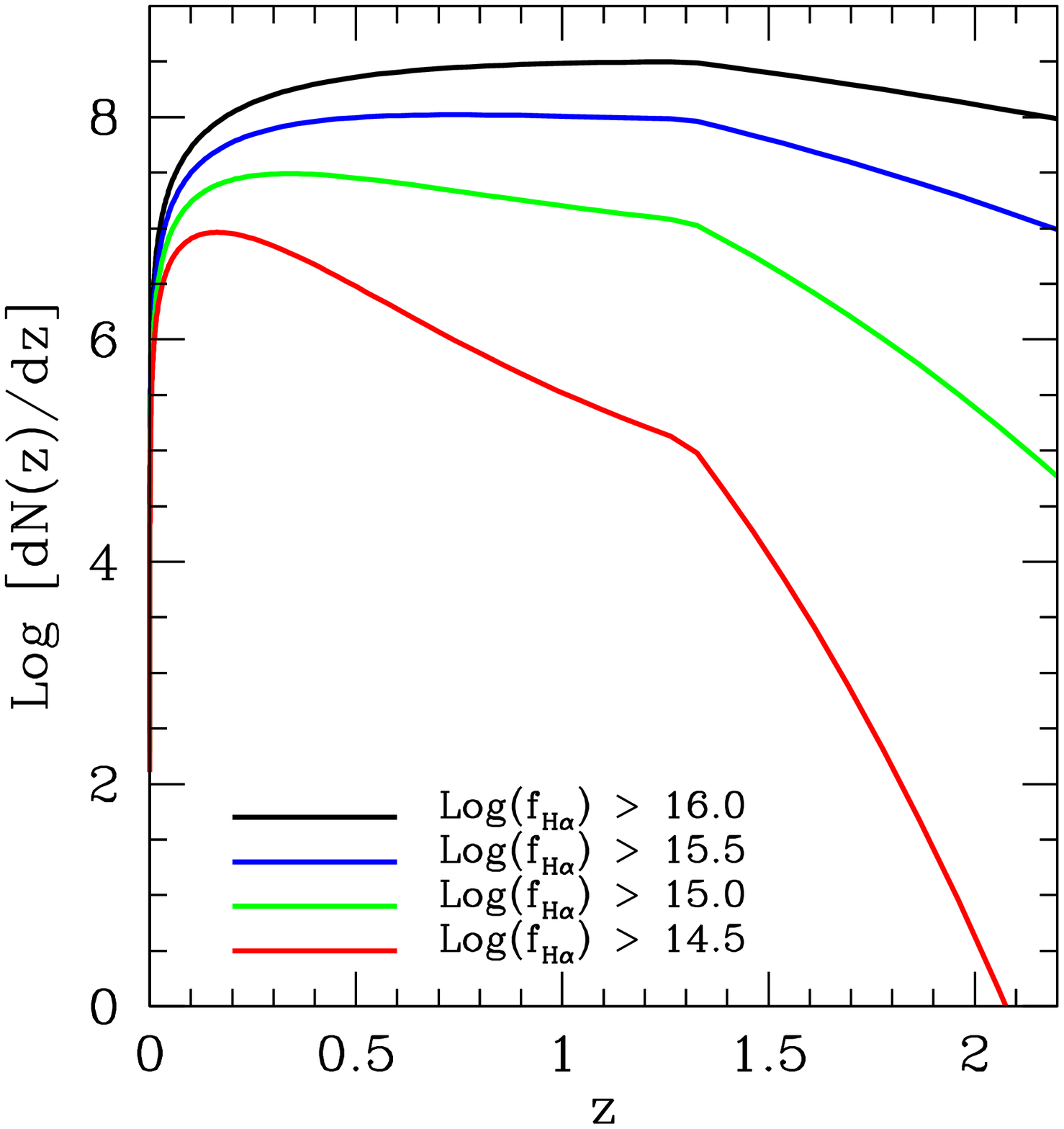}\includegraphics[width=0.5\textwidth]{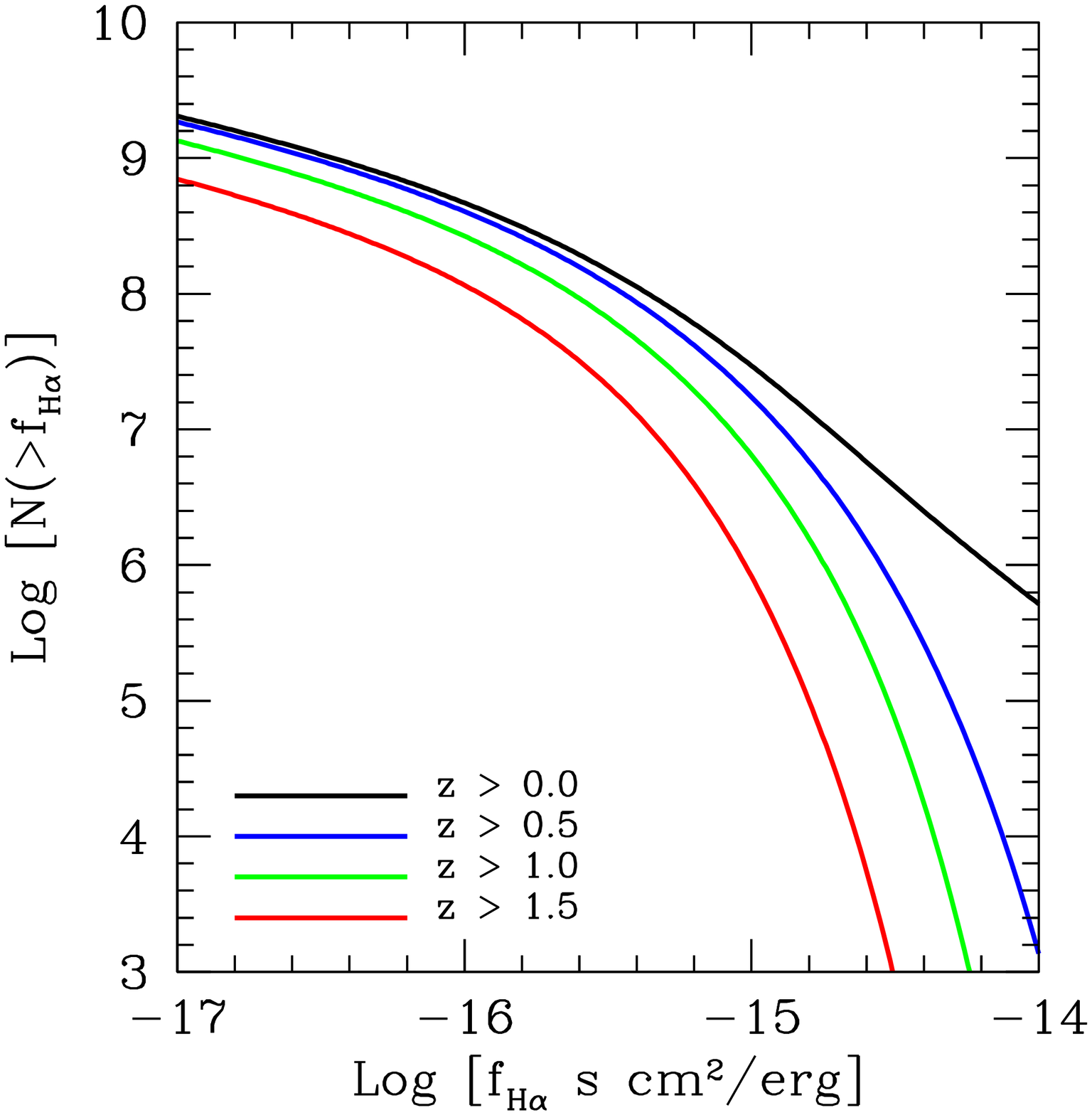}
\caption{\textit{Left panel:} Redshift distributions of H$\alpha$ emission line galaxies for four flux limits. \textit{Right panel:} Number counts of H$\alpha$ emission line galaxies versus H$\alpha$ flux for four redshift cuts.}\label{fig:Halpha}
\end{figure}

To compute the magnification bias angular power spectrum, we need to specify what the measurements of sources and lenses will be --- that is to say the specifics of the experiment(s) performing the observations. As straightforward from section~\ref{sec:magnification}, we need two populations of objects with disjoint distributions over redshift. We find that a single, deep galaxy survey performing spectroscopic measurements accurate enough to sharply separate lenses from sources would suffice. For practical purposes, we take inspiration from a Euclid-like experiment \citep{EditorialTeam:2011mu,Amendola:2012ys}. We adopt the empirical redshift distribution $\de N/\de z$ of H$\alpha$ emission line galaxies as derived by Geach et al.~\citep{2010MNRAS.402.1330G} from observed H$\alpha$ luminosity functions (see \citep{Majerotto:2012mf} for a different estimate of the H$\alpha$ emitter counts). It is presented in fig.~\ref{fig:Halpha}a, with various flux limits; note the characteristic `knee' at $z\simeq1.3$ due to redshift evolution of the luminosity function \cite[for more details, see][particularly their Eq.~(2)]{2010MNRAS.402.1330G}. Fig.~\ref{fig:Halpha}b shows the number counts of H$\alpha$ emission line galaxies versus H$\alpha$ flux, for several redshift cuts. We emphasise that fig.\ref{fig:Halpha} refers to standard \lcdm\ cosmology.

To optimise the choice of the crossover redshift $z_c$ between lenses and sources, we proceed by computing the cumulative signal-to-noise ratio
\begin{equation}
\mathrm{SNR}=\sum_\ell\frac{C^{\mu_sg_l}(\ell)}{\Delta C^{\mu_sg_l}(\ell)}\;,\label{eq:snr}
\end{equation}
where
\begin{equation}
\Delta C^{\mu_sg_l}(\ell)=\sqrt{\frac{\left[C^{\mu_sg_l}(\ell)\right]^2+\left[C_l(\ell)+1/N_l\right]\left[C_s(\ell)+1/N_s\right]}{\Delta\ell(2\ell+1)f_\mathrm{sky}}}\label{eq:error}
\end{equation}
is the statistical error on the estimated magnification bias angular power spectrum \citep{Zhang:2005eb}. Here, $C_l(\ell)\equiv C^{g_l}(\ell)$ is simply the auto-correlation angular power spectrum of foreground galaxies (the lenses), whilst $C_s(\ell)$ contains contributions from the auto-correlation of background galaxies (the sources), as well as their weak lensing convergence angular power spectrum and the cross-correlation between source number density fluctuations and weak lensing convergence, namely $C_s(\ell)=C^{g_s}(\ell)+C^{\kappa_s}(\ell)+C^{\kappa_sg_s}(\ell)$. Indeed, these last two contributions are not negligible for high-redshift objects. On the contrary, weak-lensing effects for foreground, low-redshift objects may be safely neglected. The quantities $N_l$ and $N_s$ are the lens and source number densities per steradian, respectively. Finally, $\Delta\ell$ is the width of the multipole band where the power spectrum estimator is averaged upon, and $f_\mathrm{sky}$ is the fraction of surveyed 
sky. As often done in the literature, we here adopt $\Delta\ell=1$. We also adhere to the Euclid Red Book specifications of a $15,000\,\mathrm{deg}^2$ survey area, corresponding to $f_\mathrm{sky}\simeq0.36$.

\begin{figure}
\centering
\includegraphics[width=0.75\textwidth]{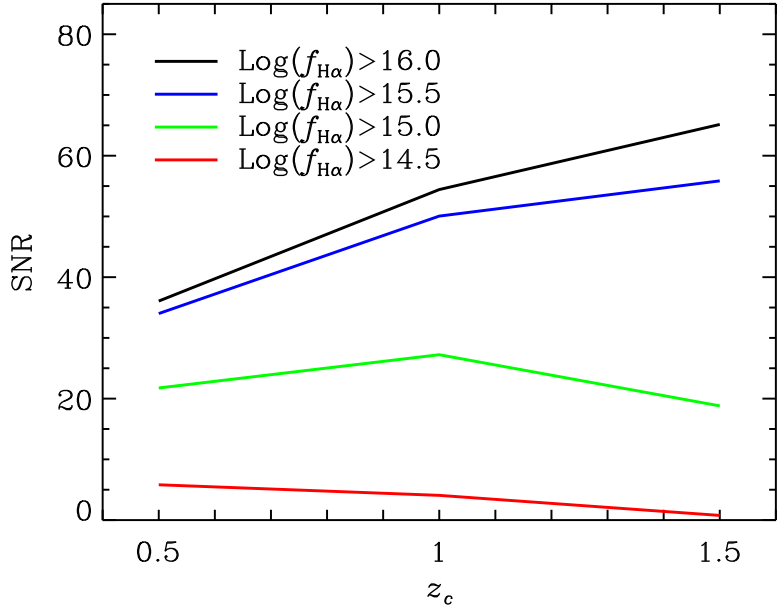}
\caption{The cumulative SNR for the magnification bias angular power spectrum as a function of the crossover redshift $z_c$ that separates `lenses' from `sources'. Results are shown for the same four different H$\alpha$ limiting fluxes shown in fig. \ref{fig:Halpha}.}\label{fig:snr}
\end{figure}

We calculate the SNR in a multipole range $\ell\in[10,\,5000]$ for a three redshift cuts and four flux thresholds --- as in fig.~\ref{fig:Halpha}, $z_c=0.5$, $1.0$, $1.5$ and $f_{\mathrm H\alpha}>10^{-16}$, $10^{-15.5}$, $10^{-15}$, $10^{-14.5}\,\mathrm{erg\,cm^{-2}\,s^{-1}}$. From fig.~\ref{fig:snr}, we see that the higher SNR is obtained with $z_c=1.5$ and $f_{\mathrm H\alpha}>10^{-16}\,\mathrm{erg\,cm^{-2}\,s^{-1}}$, that would in principle be our choice. However, the difference with respect to $f_{\mathrm H\alpha}>10^{-15.5}\,\mathrm{erg\,cm^{-2}\,s^{-1}}$ is rather small. For this reason, we choose this more conservative value as fiducial, since it is also in agreement with the $3\times10^{-16}\,\mathrm{erg\,cm^{-2}\,s^{-1}}$ cut predicted for the Euclid satellite. It implies $\am=1.64$, which means that galaxies in the background sample do get magnified.

\begin{figure}
\centering
\includegraphics[width=0.75\textwidth]{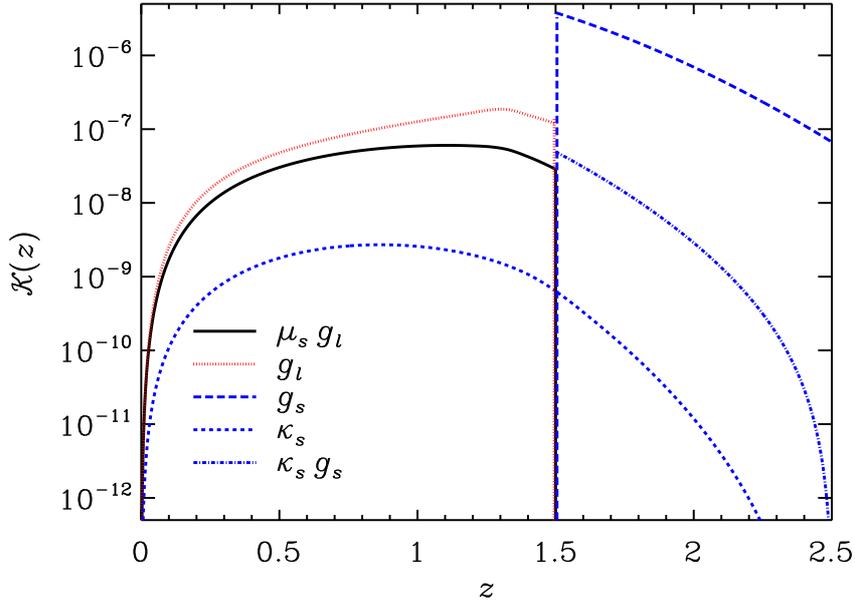}
\caption{Kernels of the considered angular power spectra: magnification bias (solid, black); auto-correlation of lens number density fluctuations (dot-dashed red); the same for sources (long-dashed blue); auto-correlation of source weak lensing convergence (short-dashed blue); and their cross-correlation (dotted blue).}\label{fig:kernel}
\end{figure}

To help in having a clearer idea of how the magnification bias behaves in the standard \lcdm\ model and for the Euclid-like experiment we considered, we depict in fig.~\ref{fig:kernel} the kernels that enter the magnification bias power spectrum and its statistical noise. They are straightforwardly defined as
\begin{equation}
\mathcal K^{XY}(z)=W^X(z)W^Y(z)\;,
\end{equation}
where $X$ and $Y$ stand for any two of the cosmological observables of interest. Specifically, we plot the kernel for the weak lensing convergence of background sources, $\mathcal K^{\kappa_s}(z)=[W^{\kappa_s}(z)]^2$ (short-dashed blue) and the kernel for the number counts of foreground lens galaxies, $\mathcal K^{g_l}(z)=[W^{g_l}(z)]^2$ (dotted red). Moreover, we show the analogous of the latter for background galaxy number counts, $\mathcal K^{g_s}(z)=[W^{g_s}(z)]^2$ (long-dashed blue), as well as their cross-correlation with weak lensing convergence, $\mathcal K^{\kappa_sg_s}(z)=W^{\kappa_s}(z)W^{g_s}(z)$ (dot-dashed blue). Finally, the solid, black curve refers to the magnification bias kernel entering eq.~\eqref{eq:C^mg}, viz.\ $\mathcal K^{\mu_sg_l}(z)=2\am W^{\kappa_s}(z)W^{g_l}(z)$. A clear feature is that all the kernels containing at least one galaxy window function have a sharp cut at $z_c=1.5$, as a consequence of the choice that we have made on the disjoint distributions of lenses and sources. 
On 
the contrary, this does not happen for the convergence kernel of source galaxies. Indeed, even though the redshift distribution of the sources, $\de N_s/\de z$, has been cut at $z_c=1.5$, its window function (eq.~\ref{eq:Ws}) does not vanish when $z<z_c$. This is of course because the high- and low-redshift galaxies alike are lensed by low-redshift structures. For this reason, there will be always a stronger signal in the high-redshift bins. This is another important fact for SNR considerations in choosing the bins \cite[e.g.][]{Hu:2000ee}. We remind the reader that all weights in angular power spectra have the meaning of probability distributions for the sources of the signal.\footnote{We are here talking about sources of \textit{signal}, not sources seen as opposite to lenses.} Therefore, they must be normalised to unity over the redshift range considered. This is why galaxy windows in fig.~\ref{fig:kernel} do not match at the crossover redshift, for the number of objects in each sample is different.

\section{Magnification bias in the presence of primordial magnetic fields}\label{sec:forecasts}
\begin{figure}
\centering
\includegraphics[width=0.75\textwidth]{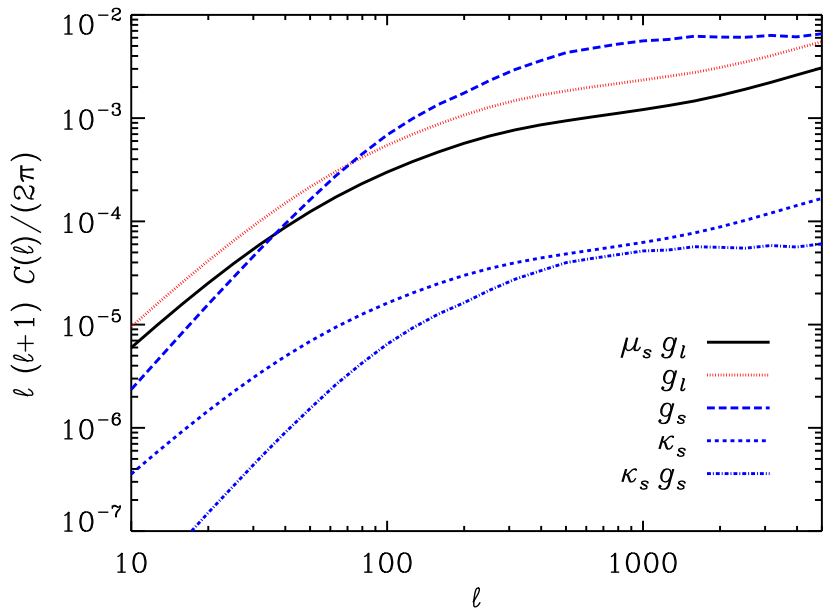}
\caption{Angular power spectra of interest for the analysis: magnification bias power spectrum, $C^{\mu_sg_l}(\ell)$ (solid, black); auto-correlation of lens number density fluctuations, $C^{g_l}(\ell)$ (dot-dashed red); the same for sources, $C^{g_s}(\ell)$ (long-dashed blue); auto-correlation of source weak lensing convergence, $C^{\kappa_s}(\ell)$ (short-dashed blue); and its cross-correlation with source clustering, $C^{\kappa_sg_s}(\ell)$ (dotted blue).}\label{fig:spectra}
\end{figure}

In fig.~\ref{fig:spectra} we illustrate all the main ingredients appearing in eq.~\eqref{eq:error} for the reference \lcdm\ model and the adopted survey design.
In detail: the solid, black curve shows the magnification bias power spectrum, $C^{\mu_sg_l}(\ell)$; the dotted red curve displays the auto-correlation of lens number density fluctuations; the same for sources is shown in long-dashed blue; the auto-correlation of their weak lensing convergence is represented in short-dashed blue; and finally, their cross-correlation, $C^{\kappa_sg_s}(\ell)$, is the dot-dashed blue curve. As expected, the magnification bias angular power spectrum lies within the galaxy spectrum of the foreground lenses and the convergence spectrum of the background sources. This happens because it is technically a cross-correlation of the two signals. Similarly, one would also expect $C^{\kappa_sg_s}(\ell)$ to lie in between its two progenitors. However, we must not forget that we here use a `maimed' galaxy redshift distribution, which by construction lacks --- in the case of sources --- low-redshift objects. If we look at the corresponding kernels in fig.~\ref{fig:kernel}, it is then easy to 
understand what happens. On the one hand, the source power spectrum $C^{g_s}(\ell)$ is substantially larger than $C^{\kappa_s}(\ell)$, because lensing is in general fainter than the clustering signal of the same sources. On the other hand, when we cross-correlate the two observables, the new kernel loses all the low-redshift information, as illustrated by the dot-dashed blue curve in fig.~\ref{fig:kernel}. Thus, even though the convergence kernel is smaller than that of the cross-correlation, when we integrate over the line of sight, the former collects more contribution from low redshifts and it turns out to provide a larger signal than $C^{\kappa_sg_s}(\ell)$. The redshift cut which affects clustering window functions is also responsible --- through Limber's approximation, which sets $k=\ell/\chi(z)$ --- for the unexpected relative behaviour of $C^{g_l}(\ell)$ and $C^{g_s}(\ell)$.

\begin{figure}
\centering
\includegraphics[width=0.51\textwidth]{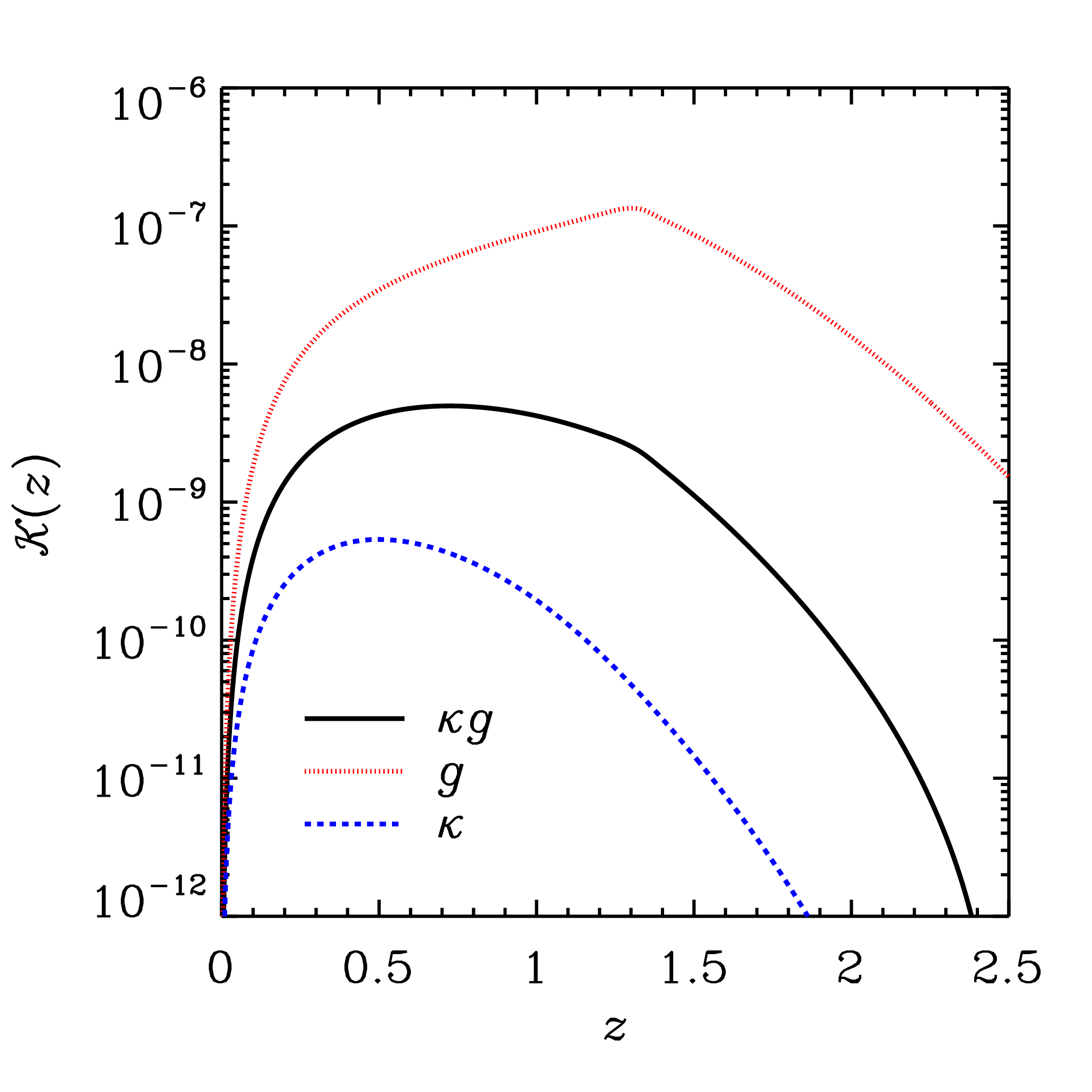}\includegraphics[width=0.51\textwidth]{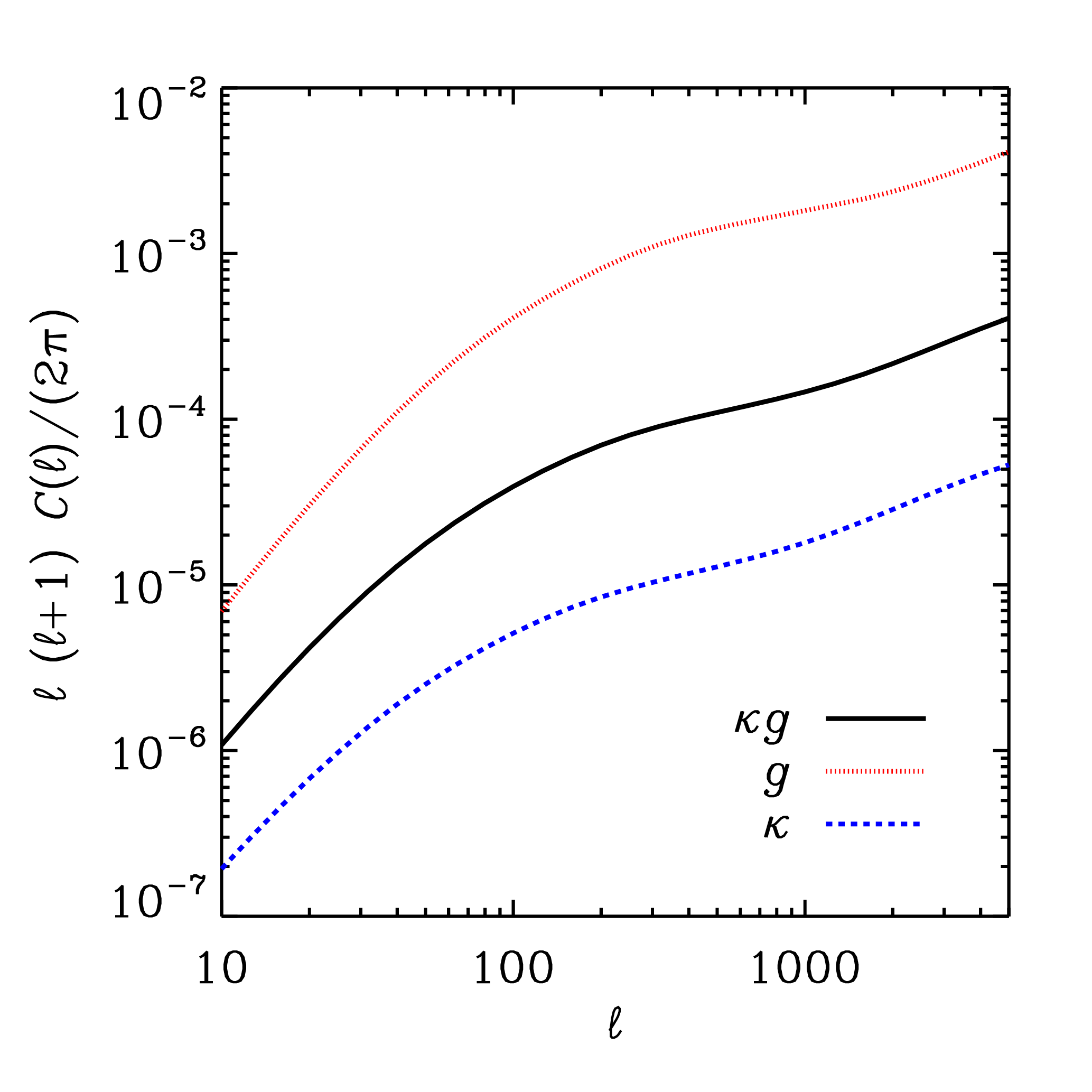}
\caption{Analogous of figs~\ref{fig:kernel}-\ref{fig:spectra} for the standard galaxy-lensing cross-correlation. \textit{Left panel:} kernels of the auto-correlations of galaxy clustering (dotted red) and weak lensing convergence (short-dashed blue). \textit{Right panel:} corresponding angular power spectra (same colour code).}\label{fig:spectra-cross}
\end{figure}

To better clarify these differences between the magnification bias effect and the usual cross-correlation between galaxy clustering and weak lensing, we suggest the reader to look at fig.~\ref{fig:spectra-cross}. In the left panel, we present the analogous of fig.~\ref{fig:kernel} for galaxy-lensing cross-correlation, whose resulting power spectra are shown in the right panel. Specifically, we compute the angular power spectra of galaxy clustering, weak-lensing convergence and their cross-correlation for a redshift distribution of sources which has not been cut anywhere. Much as expected, the cross-correlation $C^{\kappa g}(\ell)$ (solid, black) is now within the two auto-correlations, contrarily to what we have seen in the case of $C^{\kappa_sg_s}(\ell)$.

\begin{figure}
\centering
\includegraphics[width=0.5\textwidth]{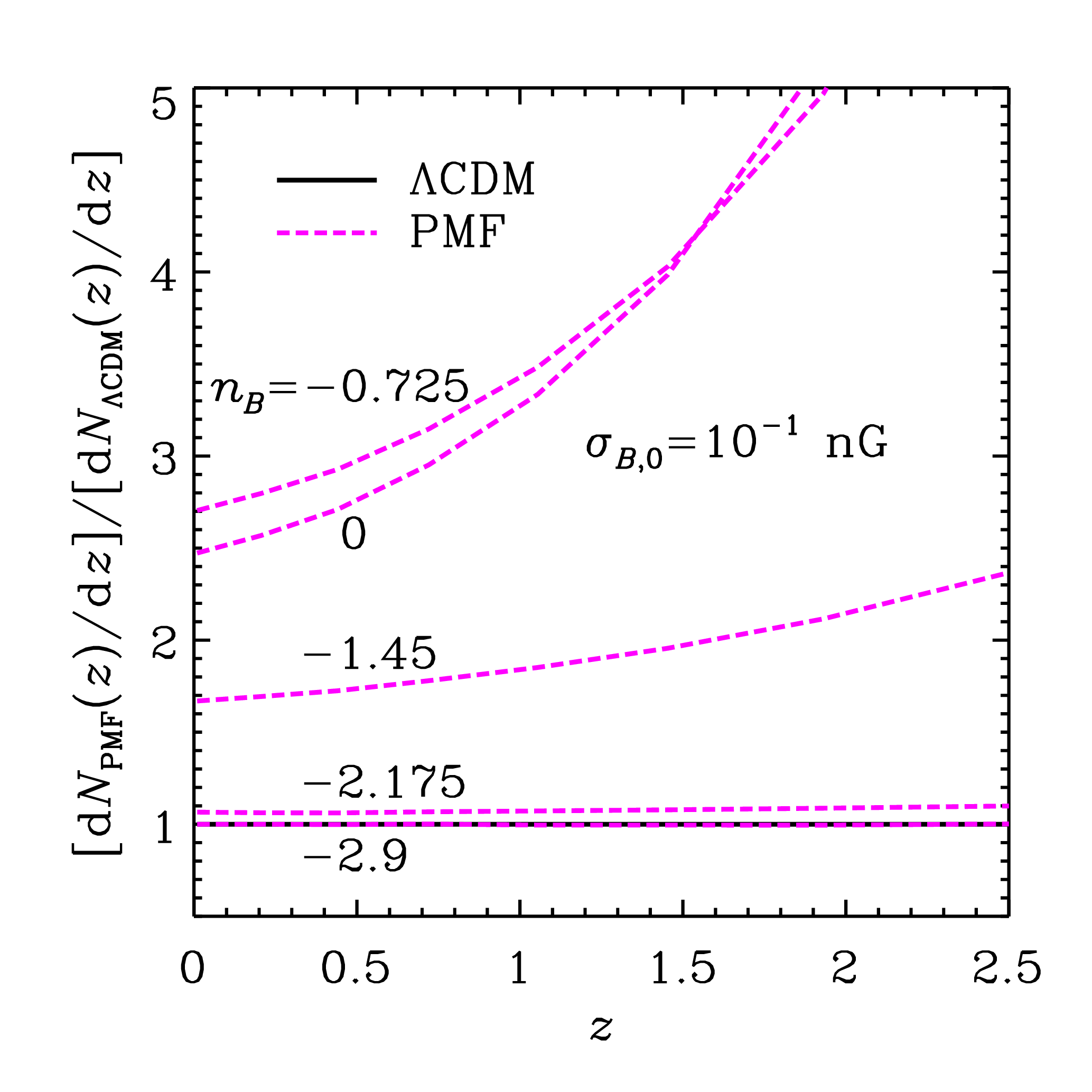}\includegraphics[width=0.5\textwidth]{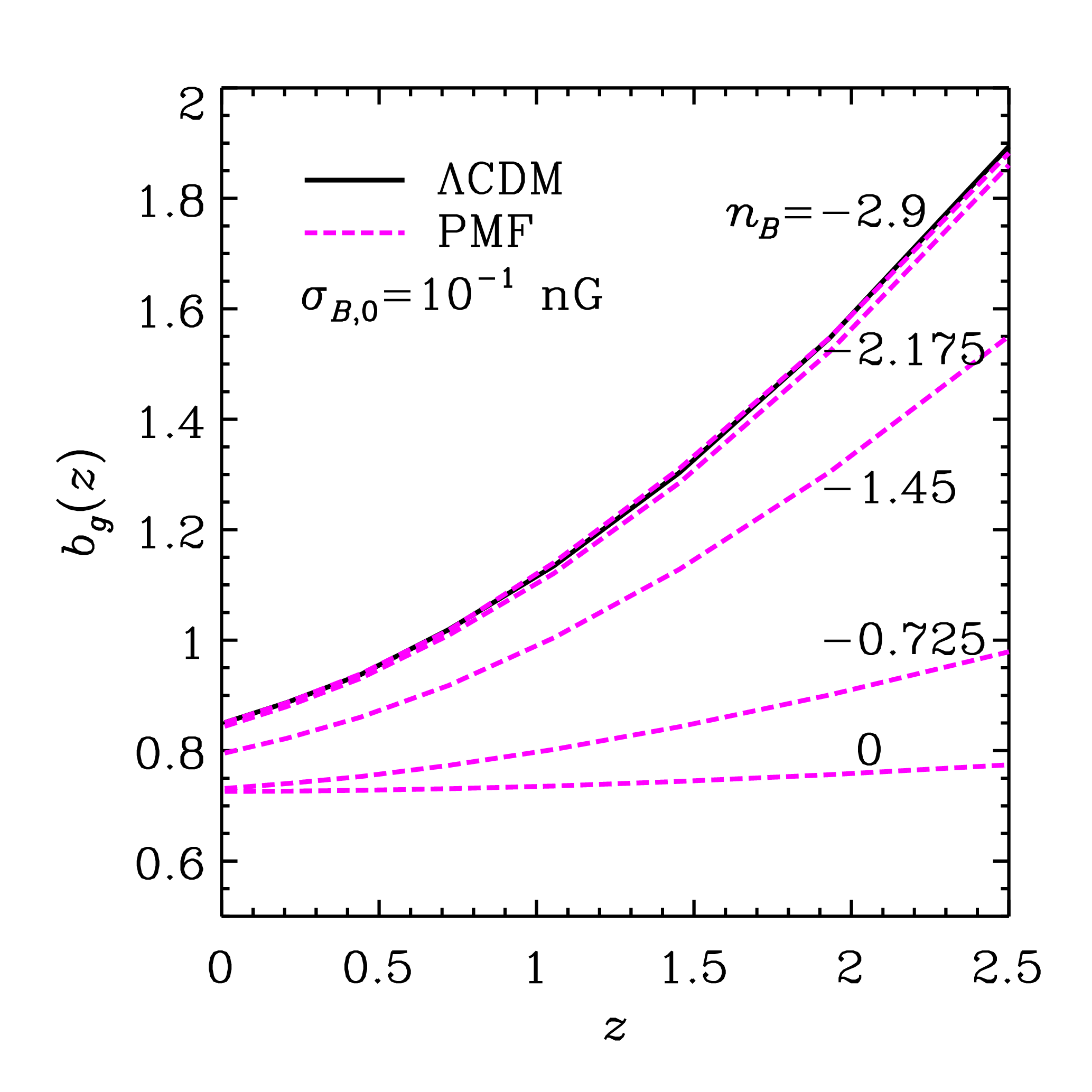}
\caption{Comparison between the reference \lcdm\ cosmology (solid, black) and models with five different PMF spectral slopes (dashed magenta), i.e.\ $n_B=-2.9$, $-2.175$, $-1.45$, $-0.725$ and $0$, with $\sbo(\lambda=1h^{-1}\,\mathrm{Mpc})=10^{-1}\,\mathrm{nG}$. \textit{Left panel:} Redshift distribution of Euclid H$\alpha$ emission line galaxies. \textit{Right panel:} H$\alpha$ galaxy bias $b_g(z)$ as a function of redshift.}\label{fig:dNdz-bias}
\end{figure}

So far, we have presented the magnification bias effect for the reference \lcdm\ cosmology, to the aim of understanding the basic physical processes at work. Now, we shall study what happens in the presence of non-vanishing early magnetism. We compute the effective bias, the galaxy redshift distributions and the non-linear matter power spectrum according to sect.~\ref{sec:clustering}. For what concerns the galaxy counts, we consider the H$\alpha$ galaxy redshift distribution $\de N/\de z$ derived by Geach et al. \citep{2010MNRAS.402.1330G} for the fiducial \lcdm\ cosmology (blue curve of fig.~\ref{fig:Halpha}a). Then, we correct it for the impact of PMFs by adopting a halo occupation distribution \citep{Fedeli:2010ud,Fedeli:2012rr}, further modifying the halo abundance according to sect.~\ref{ssec:bias}. In this way, we use an effective redshift distribution given by
\begin{equation}
\frac{\de N_\mathrm{eff}}{\de z}(z)=\frac{\de N_\mathrm{PMF}}{\de z}(z)\left[\frac{\de N_\textrm{\lcdm}}{\de z}(z)\right]^{-1}\frac{\de N}{\de z}(z)\;.\label{eq:dNdz-eff}
\end{equation}
By doing so, we properly take the deviations in the halo mass function occurring because of PMFs into account, though galaxy counts still follows H$\alpha$ emitter expectations in the fiducial \lcdm\ case. The result is presented in fig.~\ref{fig:dNdz-bias}a, for various values of the PMF spectral slope $n_B$ and for the largest considered PMF amplitude, $\sbo=10^{-1}\,\mathrm{nG}$ --- in order to enhance deviations from \lcdm. As it easy to understand, the smaller the value of $n_B$, the smaller the effects of PMFs. Indeed, the magenta dashed curve corresponding to $n_B=-2.9$ is basically on top of the \lcdm\;curve. Conversely, the stronger clustering caused by the presence of significant PMFs acts by augmenting the number of clustered haloes and, consequently, observed galaxies. This can be seen by looking at the curves with larger values of $n_B$. Besides the redshift distribution of observed galaxies, PMFs also alter the magnification bias power spectrum of eq.~\eqref{eq:C^mg} through the galaxy bias, as we discussed in sect.~\ref{ssec:bias}. The effect of primordial magnetism on H$\alpha$ galaxy bias, $b_g(z)$, is presented in fig.~\ref{fig:dNdz-bias}b, again for various values of $n_B$ and for $\sbo=10^{-1}\,\mathrm{nG}$. In this case, the larger the contribution from PMFs, the smaller the bias.

\begin{figure}
\centering
\includegraphics[width=\textwidth]{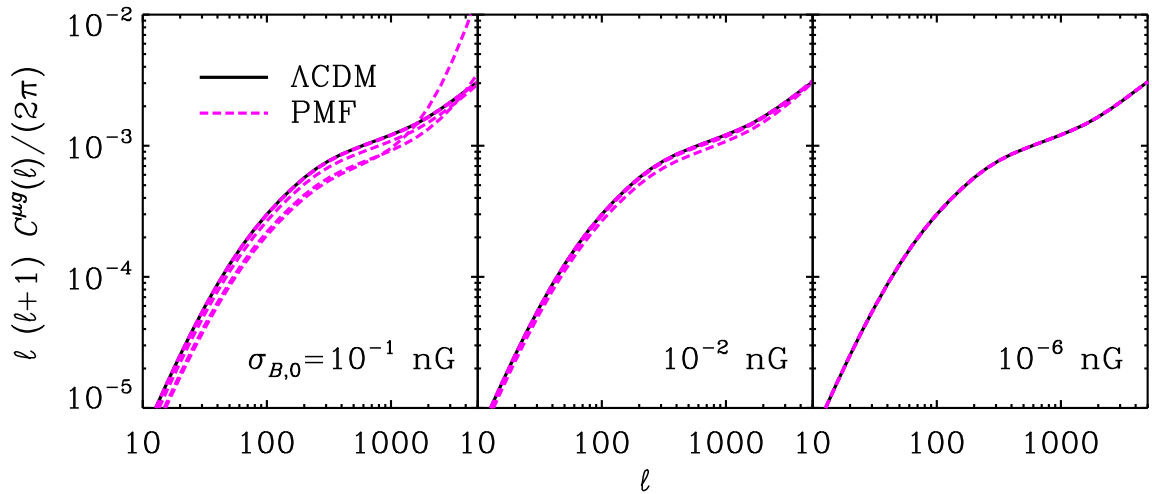}
\caption{Magnification bias angular power spectra in the presence of PMFs (dashed, magenta curves) for various magnetic spectral indices and three amplitudes, compared to the standard \lcdm\ prediction (solid, black line). Values of $n_B$ decrease from $0$ (bottom curve) to $-2.9$ (top curve).}\label{fig:spectra-PMF}
\end{figure}

Fig.~\ref{fig:spectra-PMF} finally shows the combination of all the above-described effects on the PMF magnification bias angular power spectrum, for different values of the magnetic amplitude $\sbo$ and various values of the magnetic spectral slope $n_B$. In sect.~\ref{sec:clustering}, we have shown that there are three major effects caused by the presence of PMFs. First of all, for a given PMF amplitude there is an overall augment in the clustering power as the spectral index $n_B$ increases. This causes a larger number of clustered haloes compared to standard \lcdm, which ultimately implies a larger galaxy redshift distribution. Secondly, such a larger clustering power kicks in at small scales, low wavenumbers are therefore not influenced by the PMFs. Lastly, the galaxy bias is generally lower than what expected in the reference \lcdm\ cosmology, for a fixed PMF amplitude and increasing $n_B$. Therefore, it is not a priori given what effect will dominate, and what will hence be the combined effect on the magnification bias signal. As it happens, the smaller bias induced by PMFs counteracts the larger $\de N/\de z$, and the spectra are always smaller than what obtained for \lcdm. However, such an effect is heavily dependent on the PMF amplitude, as it is clear by looking at the three panels of fig.~\ref{fig:spectra-PMF}. Indeed, for smaller amplitudes $\sbo$ the magnification bias spectrum tends to get closer to the \lcdm\;prediction than for larger amplitudes. On the other hand, the enhanced power at small physical scales which affects the three-dimensional power spectrum (see sect.~\ref{ssec:powerspectrum} and ref.~\cite[][fig.~7]{Fedeli:2012rr}) translates into a larger expected signal at large angular wavenumbers, for large values of the PMF spectral index. This last effect eventually dominates at small angular scales, and can be seen in the leftmost panel of fig.~\ref{fig:spectra-PMF}. There the bottom magenta curve, referring to $n_B=0$, is much smaller than the \lcdm\ spectrum at large angular scales, but starts increasing around $\ell\sim 1000$ and eventually overtakes it.

\subsection{Constraints on primordial magnetic fields}
To estimate the constraints on PMF amplitude that can be obtained with magnification bias from a spectroscopic galaxy survey as that which we adopted, we calculate a $\Delta\chi^2$ function as follows. We assume that the underlying magnification bias power spectrum is the one obtained in the standard model without PMFs, whilst the model power spectrum is the one obtained by including PMFs with a given amplitude and spectral index. In other words, we have
\begin{equation}
\Delta\chi^2(\sigma_{B,0},n_B)=\sum_{\ell=\ell_\mathrm{min}}^{\ell_\mathrm{max}}\left[\frac{C^{\mu_sg_l}(\sbo,n_B;\ell)-C^{\mu_sg_l}_\textrm{\lcdm}(\ell)}{\Delta C^{\mu_sg_l}_\textrm{\lcdm}(\ell)}\right]^2\;.\label{eq:chi2}
\end{equation}
We fix $\ell_\mathrm{min}=10$, not to include angular multipoles where Limber's approximation is less accurate. On the other hand, we allow for two values of the maximum allowed multipole, specifically $\ell_\mathrm{max}=1000$ and $5000$. If the first represents a fairly conservative choice, the last is for instance the official Euclid reference value \citep{EditorialTeam:2011mu}.

\begin{figure}
\centering
\includegraphics[width=0.5\textwidth]{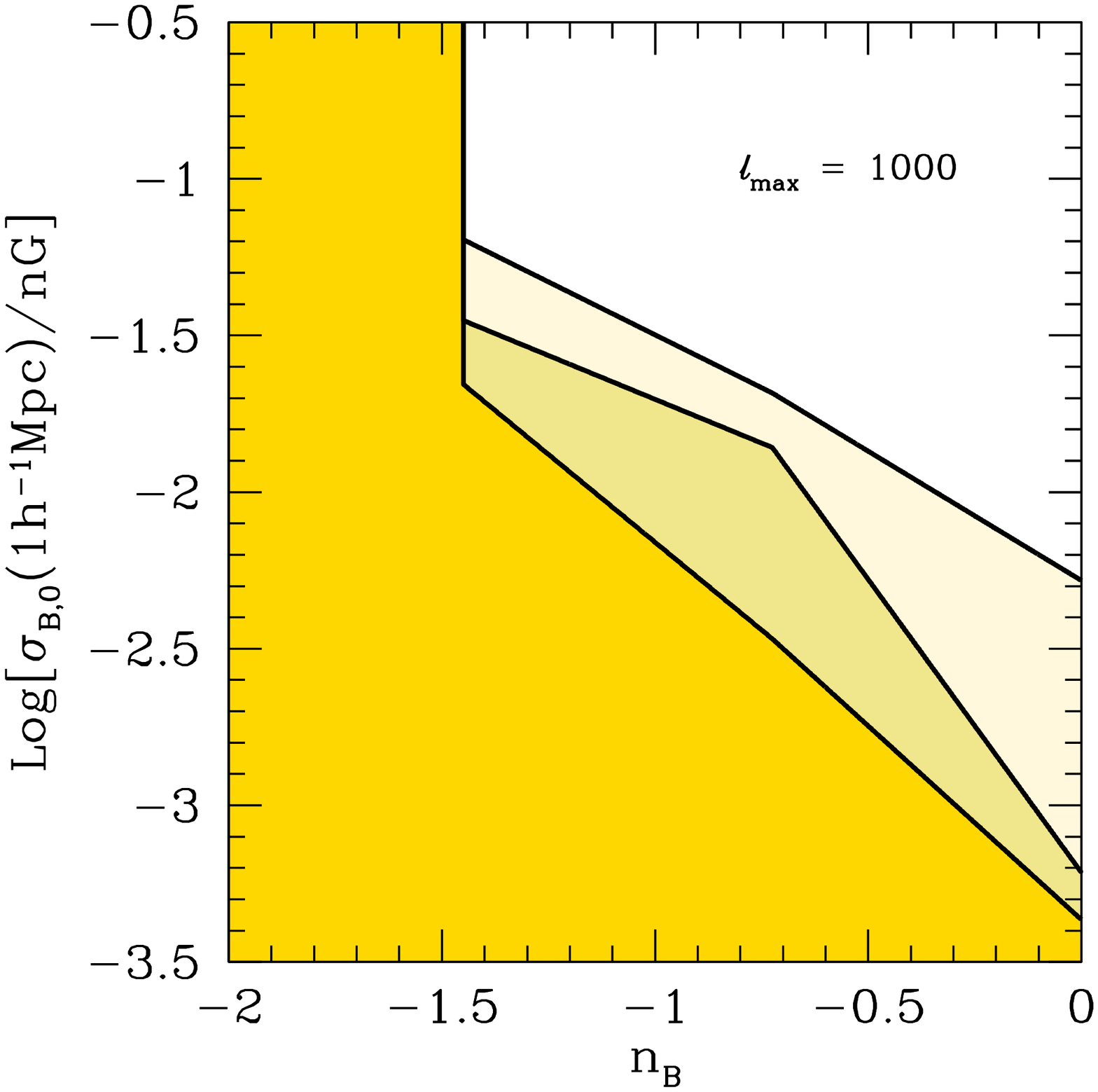}\includegraphics[width=0.5\textwidth]{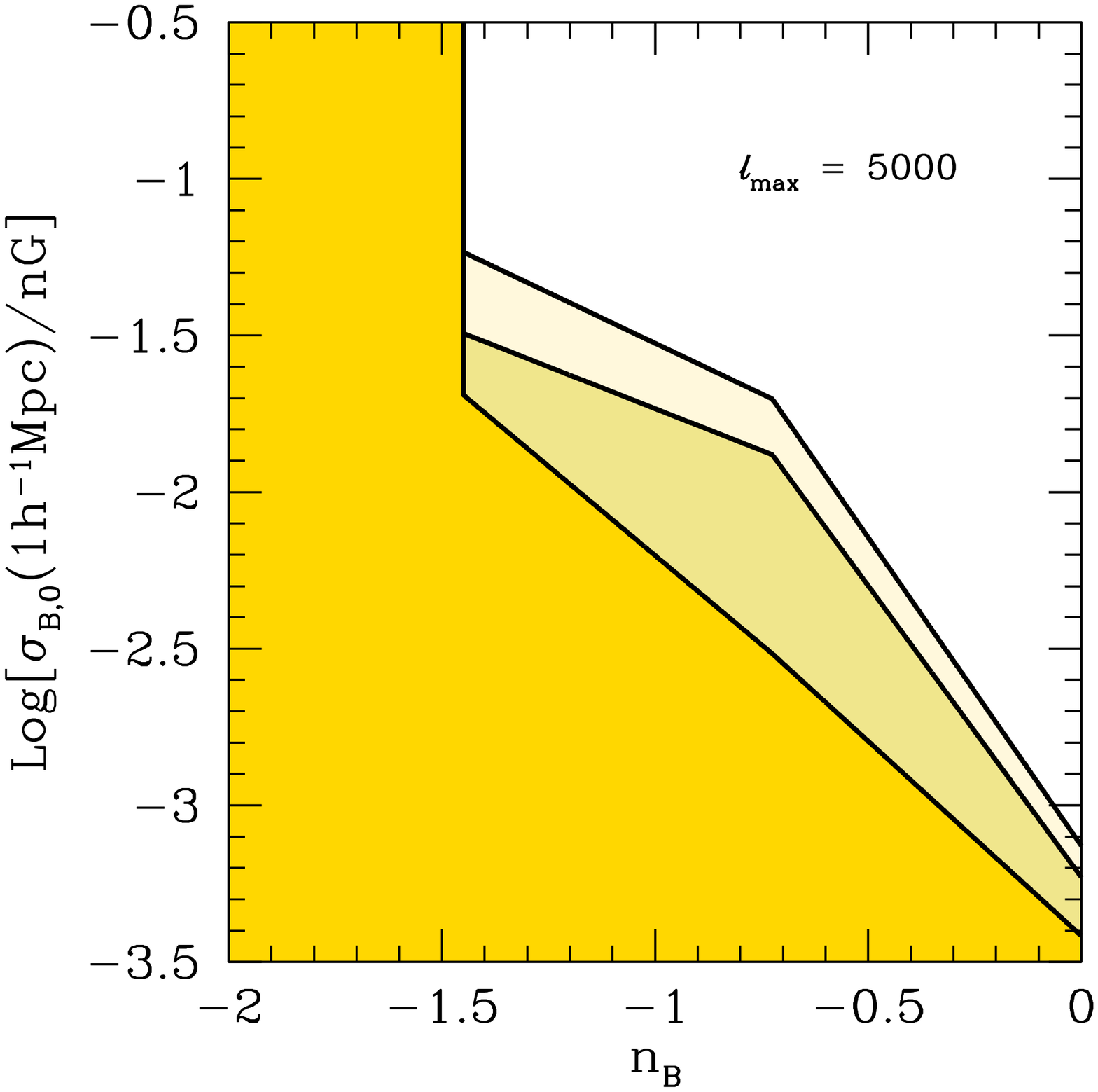}
\caption{Upper limits on the PMF amplitude as a function of the spectral index $n_B$,
obtained from magnification bias with $\ell_\mathrm{max}=1000$ and $5000$ from left to right. The different contours refer to different CLs. Specifically: $99.7\%$ CL falls within the outer, solid line; $95.4\%$ CL is the middle, light-yellow contour; and $68.3\%$ CL is the leftmost, dark-yellow area.}\label{fig:chi2}
\end{figure}

In fig.~\ref{fig:chi2} we present the Confidence Levels (CLs) for $\sbo$ as a function of $n_B$ for the $\ell_\mathrm{max}=1000$ and $5000$, from left to right. Different contours are for $99.7\%$, $95.4\%$ and $68.3\%$ CLs. Quite unexpectedly, there is not a significant difference between the two. This is because the major contribution to the $\Delta\chi^2$ function comes from the fact that there is an overall depression of the magnification bias signal in the presence of PMFs along the whole range of angular multipoles, compared to \lcdm\ (fig.~\ref{fig:spectra-PMF}). It is only at $\ell\gtrsim1000$ at most that the enhancement in the matter clustering power kicks in and causes a rise of the $C^{\mu_sg_l}(\ell)$ --- look for instance at the dashed, magenta curve referring to $n_B=0$ in fig.~\ref{fig:spectra-PMF}. However, such an increment in power actually starts from a spectrum which is smaller than that of \lcdm. That is to say, albeit the signal increases the relative difference with respect to \lcdm\ 
decreases. Then, it is only after it overtakes the \lcdm\ signal that $\Delta\chi^2$ can appreciably increase again.

The general trend of CLs as a function of magnetic spectral index is similar to what found in ref.~\cite[][fig.~9]{Fedeli:2012rr} for cosmic shear, namely the constraints become increasingly better for increasing values of $n_B$. This is due to the generic feature that the impact of PMFs on large-scale structure formation is larger for larger spectral slopes at fixed amplitude. Magnification bias alone will be able to constrain e.g.\ $n_B=0$ already for a PMF amplitude of $\sim5\times10^{-4}\,\mathrm{nG}$, whereas cosmic shear may detect it only for stronger PMFs, such as $\sigma_{B,0}\sim10^{-3}\,\mathrm{nG}$. This shows that magnification bias, for all being only a second-order effect, will be able to place competitive bounds on early magnetism. On the other hand, PMF effects on the magnification bias signal are lower than those on cosmic shear for small values of the spectral index. Below $n_B\simeq-1.5$ the constraints on the amplitude of PMFs reach $\sigma_{B,0} = 10^{-1}$ nG (the higher value we are 
considering in our calculations, hence the jump at infinity of the confidence levels), meaning that constraints for those spectral indices will not competitive with other probes.

\section{Conclusions}\label{sec:conclusions}
In this paper we have scrutinised the impact of PMFs on the magnification-induced galaxy bias. This is a second-order effect on galaxy clustering occurring because observed luminosities of distant galaxies are increased by the lensing effect due to intervening matter along the line of sight. Specifically, the `sources' are magnified in size by the convergence caused by the `lenses', whilst their surface brightness is conserved. This leads to an increase in the total observed luminosity of a source. From an observational viewpoint, there are two competing effects. On the one hand, the flux increases due to magnification of distant faint sources, thus augmenting the observations above a certain magnitude threshold. On the other hand, there is a dilution of the number density due to the stretching of the solid angle caused by lensing. Since the presence of PMFs alters both the matter power spectrum and the halo mass function --- and thus the bias and the redshift distribution of galaxies --- magnification bias 
can efficiently detect their presence.

To estimate the potential of magnification bias in detecting and constraining PMFs, we have chosen a spectroscopic galaxy survey, in a way similar to the forthcoming ESA Euclid satellite. As a working hypothesis, we have adopted the predicted galaxy distribution of sources and number counts as a function of H$\alpha$ flux (fig.~\ref{fig:Halpha}) of ref.~\citep{2010MNRAS.402.1330G}, and modified it to take PMF presence into account (fig.~\ref{fig:dNdz-bias}). Since it is necessary to have two disjoint distributions of galaxies to measure the magnification bias effect, that is to say a source and a lens sample, we have optimised the survey constraining power to the best achievable SNR (fig.~\ref{fig:snr}). It implies a separation between sources and lenses at $z_c=1.5$, for a conservative flux limit of $10^{-15.5}\,\mathrm{erg\,cm^{-2}\,s^{-1}}$.

Summarising, we find that:
\begin{itemize}
\item In agreement with previous results, the larger the PMF spectral index $n_B$, the larger the deviations with respect to what predicted by the standard \lcdm\ model.
\item The enhancement of the clustering power at small physical scales induced by PMFs translates into a similar enhancement at large angular multipoles for the magnification bias power spectrum.
\item However, quite a different behavior appears in comparison to what found for weak-lensing cosmic shear by ref.~\citep{Fedeli:2012rr}. Indeed, being the magnification bias dependent upon the bias of lenses, it is affected by the fact that galaxy bias is smaller for more intense PMFs, even though the matter power spectrum gets enhanced. Such a combination of counteracting effects ultimately yields what we have shown in fig.~\ref{fig:spectra-PMF}. In practice, the PMF magnification bias power spectrum is in general depressed with respect to that in \lcdm, and it is only at very small angular scales that the boosted matter power spectrum beats the effect of a smaller bias and eventually overtakes the \lcdm\ signal.
\item Such rescaling of the magnification bias signal can be in principle degenerate with the normalisation of the present-day matter power spectrum. Therefore, it is important to use, alongside magnification bias, other observables like galaxy clustering and cosmic shear, in order to lift this degeneracy.
\item Thanks to its peculiar behaviour, magnification bias provides us with information which is additional to that of galaxy clustering and weak lensing sole. Indeed, a spectroscopic Euclid-like survey will be able to constrain e.g.\ $n_B=0$ already for a PMF amplitude of $\sim5\times10^{-4}\,\mathrm{nG}$, whereas cosmic shear may detect it only for stronger PMFs.
\end{itemize}

For the aforementioned reasons, it will be interesting in future work to test the combined effect of galaxy clustering, weak lensing cosmic shear and magnification bias altogether in a self-consistent way. Moreover, it will also be worth to better investigate the degeneracies that may possibly arise with other cosmological parameters or even extensions of the concordance cosmological model such as the running of the spectral index or the presence of mildly warm dark matter. However, this goes beyond the scope of the present work.

\acknowledgments We thank the anonymous referee, who helped in enhancing the quality of our results with her/his accurate reviewing. We also thank Gigi Guzzo and Eiichiro Komatsu for their comments and suggestions and M\'ario G. Santos for useful discussions at the beginning of this project. SC is funded by FCT-Portugal under Post-Doctoral Grant No. SFRH/BPD/80274/2011. CF was partially supported by the University of Florida through the Theoretical Astrophysics Fellowship, and has received funding from the European Commission Seventh Framework Programme (FP7/2007-2013) under grant agreement n$^\circ$ 267251. LM acknowledges financial contributions from contracts ASI/INAF I/023/12/0, by the PRIN MIUR 2010-2011 ``The dark Universe and the cosmic evolution of baryons: from current surveys to Euclid'' and by the PRIN INAF 2012 ``The Universe in the box: multiscale simulations of cosmic structure''.

\bibliographystyle{JHEP}
\bibliography{/home/stefano/Documents/LaTeX/Bibliography}

\end{document}